\documentstyle[12pt,epsf]{article} 
\newcommand{\VV}{{\cal V}}

\newcommand{\wt}{\widetilde}
\newcommand{\wh}{\widehat}

\newcommand{\be}{\begin{equation}}
\newcommand{\ee}{\end{equation}}
\newcommand{\ben}{\begin{eqnarray}\displaystyle}
\newcommand{\een}{\end{eqnarray}}
\newcommand{\refb}[1]{(\ref{#1})}
\newcommand{\p}{\partial}
\newcommand{\sectiono}[1]{\section{#1}\setcounter{equation}{0}}
\renewcommand{\theequation}{\thesection.\arabic{equation}}

\catcode`\@=11
\def\marginnote#1{}
\newcount\hour
\newcount\minute
\newtoks\amorpm
\hour=\time\divide\hour by60
\minute=\time{\multiply\hour by60 \global\advance\minute by-\hour}
\edef\standardtime{{\ifnum\hour<12 \global\amorpm={am}%
        \else\global\amorpm={pm}\advance\hour by-12 \fi
        \ifnum\hour=0 \hour=12 \fi
        \number\hour:\ifnum\minute<10 0\fi\number\minute\the\amorpm}}
\edef\militarytime{\number\hour:\ifnum\minute<10 0\fi\number\minute}
\def\draftlabel#1{{\@bsphack\if@filesw {\let\thepage\relax
   \xdef\@gtempa{\write\@auxout{\string
      \newlabel{#1}{{\@currentlabel}{\thepage}}}}}\@gtempa
   \if@nobreak \ifvmode\nobreak\fi\fi\fi\@esphack}
        \gdef\@eqnlabel{#1}}
\def\@eqnlabel{}
\def\@vacuum{}
\def\draftmarginnote#1{\marginpar{\raggedright\scriptsize\tt#1}}
\def\draft{\oddsidemargin -.5truein
        \def\@oddfoot{\sl preliminary draft \hfil
        \rm\thepage\hfil\sl\today\quad\militarytime}
        \let\@evenfoot\@oddfoot \overfullrule 3pt
        \let\label=\draftlabel
        \let\marginnote=\draftmarginnote
   \def\@eqnnum{(\theequation)\rlap{\kern\marginparsep\tt\@eqnlabel}%
\global\let\@eqnlabel\@vacuum}  }

\def\preprint{\twocolumn\sloppy\flushbottom\parindent 1em
        \leftmargini 2em\leftmarginv .5em\leftmarginvi .5em
        \oddsidemargin -.5in    \evensidemargin -.5in
        \columnsep 15mm \footheight 0pt
        \textwidth 250mmin      \topmargin  -.4in
        \headheight 12pt \topskip .4in
        \textheight 175mm
        \footskip 0pt
        \def\@oddhead{\thepage\hfil\addtocounter{page}{1}\thepage}
        \let\@evenhead\@oddhead \def\@oddfoot{} \def\@evenfoot{} }

\def\titlepage{\@restonecolfalse\if@twocolumn\@restonecoltrue\onecolumn
     \else \newpage \fi \thispagestyle{empty}\c@page\z@
        \def\thefootnote{\fnsymbol{footnote}} }

\def\endtitlepage{\if@restonecol\twocolumn \else  \fi
        \def\thefootnote{\arabic{footnote}}
        \setcounter{footnote}{0}}  

\catcode`@=12
\relax

\def\bea{\begin{array}}
\def\bem{\begin{displaymath}}
\def\beq{\begin{equation}}

\def\eea{\end{array}}
\def\eem{\end{displaymath}}
\def\eeq{\end{equation}}

\def\s2w{\sin^2 \theta_W}

\newcommand{\real} {{{\rm I} \kern -0.2em {\rm R}}}
\newcommand{\complex} {{{\sf I} \kern -0.48em {\rm C}}}
\newcommand{\naturel} {{{\rm I}  \kern -0.18em {\rm N}}}
\newcommand{\integer} {{{\rm Z} \kern -0.31em {\rm  Z}}}
\newcommand{\smallinteger} {{{\rm Z} \kern -0.25em {\rm  Z}}}

\relax
\relax

\begin{document}
\setcounter{footnote}{0}
\setcounter{page}{1}

{}~
\hfill\vbox{\hbox{hep-th/0009038}\hbox{MRI-P-990902}
}\break

\vskip 2.0cm

\centerline{\large \bf Some Issues in Non-commutative Tachyon
Condensation}

\vspace*{6.0ex}

\centerline{\large \rm Ashoke Sen}

\vspace*{6.5ex}

\centerline{\large \it Mehta Research Institute of Mathematics}

\centerline{\large\it
and Mathematical Physics, Chhatnag Road,}

\centerline{\large \it   Jhoosi,
Allahabad 211019, INDIA}
\vspace*{1ex}
\centerline{E-mail: asen@thwgs.cern.ch, sen@mri.ernet.in}

\vspace*{4.5ex}

\centerline {\bf Abstract}
\bigskip
Techniques of non-commutative field theories have proven to be useful in
describing D-branes as tachyonic solitons in open string theory. However,
this procedure also leads to unwanted degeneracy of solutions not present
in the spectrum of D-branes in string theories. In this paper we explore
the possibility that this apparent multiplicity of solutions is due to
the wrong choice of variables in describing the solutions, and that with
the correct choice of variables the unwanted degeneracy disappears. 

\vfill \eject

\baselineskip=18pt

\tableofcontents

\sectiono{Introduction and Summary} \label{s0}

It has been conjectured that the tachyonic vacuum in open bosonic string
theory on a D-brane describes the closed string vacuum without D-branes,
and that various soliton solutions in this theory describe D-branes of
lower dimension\cite{9902105}. Similar conjectures have also been put
forward for
superstring theories\cite{ORIGINAL,9810188,9812135}. Evidence for these
conjectures come from both, first\cite{RECK,FIRST}
and second\cite{SECOND} quantized string theories.

Recently it has been realized that the study of these conjectures can
be simplified by examining D-branes in the background of anti-symmetric
tensor fields, or equivalently, in the presence of background magnetic
field on the D-brane world-volume. In this case the world-volume 
theory on the D-brane can be described by a
non-commutative field theory\cite{9711162,9711165,9903205,9908142}. 
In particular, the
non-commutative solitons discussed in
\cite{0003160} can be used to construct exact solutions of the field
equations in the limit of infinite background magnetic
field\cite{0005006,0005031,0006071,0007078,0008023,0008064,0008214}, which
can
then be
identified to lower
dimensional D-branes. This reproduces the correct tension of the lower
dimensional D-branes, and also reproduces many of the known features of
the world-volume theory of the D-branes.

One of the shortcomings of the analysis of refs.\cite{0005006,0005031} is
that it requires infinite background magnetic field, whereas the
conjectures involving tachyon condensation are expected to hold for
arbitrary values of the background magnetic field including zero
background magnetic field. In a recent paper\cite{0007226} (see also
\cite{0007217}) it was
suggested that even if we start with zero or finite magnetic field
background, in
the tachyonic vacuum the magnetic field dynamically rolls down to infinite
value.  This proposal, if correct, would provide an exact description of
tachyonic soliton solution in all cases. However, as was found in
ref.\cite{0007226}, this proposal suffers from the problem that besides
the soliton solutions representing lower dimensional D-branes, there are
many other degenerate solutions which do not have any obvious physical 
interpretation.
Furthermore there is no obvious dynamical mechanism which makes the
magnetic field roll down to infinity, since different vacua labelled by
different values of the magnetic field all appear to have the same energy 
density.
Additional degeneracies in the set of solutions was discussed in
ref.\cite{0008013}.

In fact, the problem of unwanted degeneracy appears even in the
absence of any background field strength. To see this,
let us
consider a D-$p$ brane in the bosonic string theory in 26 dimensional flat
space-time. Before tachyon condensation, the world-volume theory of the
D-$p$ brane contains $(25-p)$ massless scalar fields representing
coordinates of the brane transverse to its world-volume. These fields can
be regarded as the Goldstone modes associated with
spontaneously broken translational symmetry in the $(25-p)$ transverse
directions. Since the tachyonic vacuum is conjectured to represent the
closed string vacuum without any D-brane, we should expect that
full (25+1) dimensional translational symmetry is restored in this vacuum. 
Hence the Goldstone modes, and consequently the flat directions in the
potential in the D-brane world-volume theory, should disappear in this
vacuum. However, in terms of the variables used in describing the
Born-Infeld action, the flat directions in the potential continue to
persist in this vacuum, thereby giving us unwanted degeneracy of the
vacuum. 

In this note we propose a resolution of the above problem, as well
as the degeneracy problem encountered in ref.\cite{0007226}. We propose
that the apparent
multiplicity of the
ground state, labelled by different values of the background magnetic
field and/or different positions of the initial D-brane, is due to the
fact that the original choice of field variables,
used in describing the Born-Infeld action on the D-brane, becomes singular
at the tachyonic vacuum, and that with the correct choice of field
variables, all these apparently different vacua correspond to the same
point in the configuration space. Besides removing the vacuum
degeneracy, this proposal also resolves the problem encountered in
ref.\cite{0007226} of unwanted soliton solutions, since
many apparantly different solutions in the
original variables represent the same field configuration in the new set
of variables.

It is best to illustrate this by drawing an analogy. Consider, for example
a particle moving in three dimensions under the influence of a potential
which has a unique minimum at the origin, and let us suppose further that
the
potential is symmetric under a rotation about the $z$-axis. In this case,
if we use the spherical polar coordinates
$(r,\theta,\phi)$ to describe the motion of the particle, then it
will appear that there are infinite number of degenerate ground states
of the system, corresponding to $r=0$, $(\theta,\phi)$ arbitrary.
Furthermore, it will not be manifest that the ground state is invariant
under the rotational symmetry about the $z$-axis; since under this
transformation $\phi$ will transform to $\phi+a$ for some constant $a$.
Only after going to a non-singular coordinate system ({\it e.g.} the
cartesian coordinate system) we see that the ground state is unique
and that it is invariant under rotation about the $z$-axis.

We argue that the case at hand, $-$ the dynamics of the D-brane around the
tachyonic vacuum $-$ is similar to the example discussed above; and that
the usual variables appearing in the (non-commutative) Born-Infeld action
correspond to a singular
coordinate system around the tachyonic vacuum.   The role of the
angular variables $(\theta,\phi)$ is played by the background gauge and
massless scalar
fields, and the role of the radial variable $r$ is played by $T-T_{min}$,
where $T$ denotes the tachyon field and $T_{min}$ is its vacuum value
where the potential has a local minimum.  Thus at $T=T_{min}$, different
values of background gauge and massless scalar fields describe the same
configuration.
In the case of the three
dimensional particle, the singularity of the spherical polar coordinate
system near the origin becomes apparent
if we examine the kinetic term of the particle in this coordinate system;
as we shall see, the same situation holds
in the case of D-branes.  (Unfortunately, however, here not all the
relevant terms are known which allows us to determine precisely the right
choice of coordinate system around the new vacuum.) 

We can push the analogy a bit further by comparing deformation of the
azimuthal angle $\phi$ to the deformation of field configurations
on the D-brane generated by various symmetry transformations, and the
deformation of the
polar angle $\theta$ to deformation of background magnetic field
strength on the D-brane.
The analog of rotational
symmetry restoration at $r=0$ will then be the restoration of translation
symmetry in directions transverse to the D-brane, and also the restoration
of the $U(\infty)$ gauge symmetry discussed in
refs.\cite{0005031,0007226} in the tachyonic vacuum.
For a non-central potential which is invariant under rotation about the
$z$-axis, different values of
$\theta$ describe inequivalent configurations, but at $r=0$ they all
correspond to the same configuration. Similarly, for the D-brane system,
we shall argue that although different background magnetic field
configurations describe inequivalent configurations away from the
tachyonic
vacuum, at the tachyonic vacuum they all correspond to the same
configuration. Finally, just as in the case of a three dimensional point
particle the rotational symmetry about the $z$-axis is restored for
general $r$ at special values of $\theta$, namely 0 and $\pi$, so in the
case of a D-brane the $U(\infty)$ symmetry is restored for a general
space-time independent tachyonic background
at special value of the magnetic field strength, namely
$\infty$\cite{0005031,0007226}.

Incidentally, we would like to mention here that string field
theory\cite{SFT}, whose variables are related to the ones appearing in the
Born-Infeld action via a nontrivial field
redefinition\cite{9912243,0005085},
automatically chooses the right set of field variables around the
tachyonic vacuum. This is seen by noting the absence of Goldstone modes,
{\it i.e.} flat directions of the potential around the new
vacuum\cite{0007153,0008033,0008127}. Indeed,
most of the hard evidence for the proposal put forward in this paper comes
from this result in string field theory.

Besides the existence of unwanted solutions, another problem that the
results of refs.\cite{0005031,0007226} suffered from was the existence of
unwanted open
string states on the world-volume of the tachyonic soliton. When we
interpret the tachyonic soliton as a D-brane, then these states
correspond to open string states with one end on the D-brane represented
by the soliton, and the other end in the vacuum. Since open strings cannot
end on the vacuum, the spectrum {\it should not} contain such states;
hence existence of these states on the soliton world-volume poses a
problem. We propose
a resolution of this problem by noting that the equations of motion of
U(1) gauge field living on the brane before tachyon condensation will
force the currents associated with such states to
vanish, and hence these states cannot exist in isolation. (This follows
the suggestion in ref.\cite{9909062}, alternative but similar proposals
have been
made in refs.\cite{9901159,0002223}.) 

The rest of the paper is organised as follows. In section \ref{s2} we
discuss how translation invariance in directions transverse to the brane
is restored at the tachyonic vacuum. In section \ref{s3} we show how the
$U(\infty)$ gauge invariance can be restored at the tachyonic vacuum even
when the background magnetic field is finite. The field redefinition
required for achieving this also removes the unwanted degeneracy of vacuum
and
soliton solutions. In section \ref{s4} we discuss how the unwanted states
living on the soliton world-volume, corresponding to open strings with one
end on the soliton and the other end in the tachyonic vacuum, might be
removed from the spectrum.

We end this section by noting that although we shall carry out our
discussion in the
context of D-branes in bosonic string theory, we expect that an identical
analysis can be
carried out for D-branes in superstring theory as well. In particular,
restoration of the full supersymmetry at the tachyonic vacuum should
follow in a manner analogous to the restoration of the translation
invariance discussed in section \ref{s2}.

\sectiono{Restoration of Translation Invariance at the Tachyonic Vacuum}
\label{s2}

Let us consider a D-$p$ brane in 26 dimensional bosonic string theory in
flat Minkowski space-time. This system has $(25-p)$ massless scalar fields
living on its world-volume, describing transverse motion of the
D-$p$-brane. These can be regarded as the Goldstone bosons associated with
the spontaneously broken translational invariance along these $(25-p)$
directions in the presence of the brane. Now when the tachyon condenses
into its ground state, we expect that the translational symmetry of the
configuration should be restored fully, and hence these Goldstone modes
should disappear. The question is: can we see this in the effective field
theory describing the system?

Let us consider the Born-Infeld action describing the system in the static
gauge\cite{9909062,WORLD}:
\be \label{e2.1}
S_{BI} \propto \int d^{p+1}x \, \VV(T)\sqrt{\det\Big(\eta_{\mu\nu} +
\p_\mu
\chi^i\p_\nu\chi^i\Big)} + \ldots\, ,
\ee
where $\chi^i$ denote the transverse coordinates of the brane $(p+1\le
i\le
25)$, $x^\mu$ denote the world-volume coordinates of the brane
$(0\le\mu\le p)$, $T$ denotes the tachyon field, and $\VV(T)$ denotes the
tachyon potential with the D-brane tension term included, so that $\VV(T)$
vanishes at the tachyonic vacuum $T=T_{min}$. For simplicity we have
ignored
the gauge field terms in \refb{e2.1}. $\ldots$ denotes terms containing 
derivatives of $T$ and $\p_\mu\chi^i$. The action \refb{e2.1} has manifest
translational invariance along the transverse directions:
\be \label{e2.2}
\delta \chi^i = a^i\, .
\ee
Note, however that this symmetry is spontaneously broken, as is indicated
by the presence of field independent terms on the right hand side of
eq.\refb{e2.2}. These do not vanish even when all the fields are set to
zero.

Now consider the tachyonic vacuum $T=T_{min}$. The right hand side of
eq.\refb{e2.2} is independent of $T$, and so continues to be
non-zero even in this vacuum. Thus it would seem that the vacuum
is still infinitely degenerate, labelled by different values of
$\langle\chi^i\rangle$, and that the $\chi^i$'s continue to represent the
Goldstone modes associated with broken translation invariance. However,
since $\VV(T_{min})=0$, the kinetic term of $\chi^i$ in \refb{e2.1} vanishes
at $T=T_{min}$.
Thus
the fields $\chi^i$ are not good coordinates for describing the
field configuration around the vacuum $T=T_{min}$. Unfortunately, due to 
our limited
knowledge of the structure of the effective action around the
tachyonic vacuum, we are unable to
determine the precise expressions for the `good' variables in terms of $T$
and
$\chi^i$.
However, since the kinetic term of 
$\chi^i$ vanishes at $T=T_{min}$, we
would
expect that the good choice of field variables, $\wt T$ and $\wt\chi^i$,
should be
related to $T$ and $\chi^i$ by functional relations of the form:
\be \label{e2.3a}
\wt\chi^i = h^i(T, \vec \chi)\, ,\qquad \wt T = h(T, \vec \chi)\, ,
\ee
with $h^i(T,\vec\chi)$ and $h(T, \vec\chi)$ having the property
that\footnote{For our purpose it will be sufficient if these relations
hold for constant $\vec \chi$ configurations.} 
\be \label{e2.3b}
h^i(T_{min},\vec\chi) = 0\, , \qquad h(T_{min}, \vec\chi) 
= \wt T_{min}\, .
\ee
Here $\wt T_{min}$ is a constant which can be set to zero by suitably
redefining $\wt T$.
In these variables
eq.\refb{e2.2} takes the form:
\be \label{e2.5}
\delta\wt\chi^i = {\p h^i(T,\vec\chi)\over \p \chi^j} a^j\, ,
\qquad \delta\wt T = {\p h(T,\vec\chi)\over \p \chi^j} a^j\, .
\ee
Since eq.\refb{e2.3b} is satisfied for all $\vec\chi$, we see that
$\delta\wt\chi^i$ and $\delta\wt T$ vanish at $T=T_{min}$. Thus translation
invariance along directions transverse to the D-brane is restored around
this vacuum.

As has already been noted in the introduction, string field theory
automatically chooses the right coordinate system around the vacuum
$T=T_{min}$, since the string field theory 
potential expanded around $T=T_{min}$
has no flat direction\cite{0007153,0008033}. Indeed, this result of string
field theory can be
turned around to conclude that the correct choice of coordinate system
around $T=T_{min}$ must be of the form given in eqs.\refb{e2.3a},
\refb{e2.3b}
so that in this coordinate system there are no flat directions of the
potential.

\sectiono{Lifting of Degeneracy of Solutions} \label{s3}

In the analysis of ref.\cite{0007226} the authors proposed a specific form
of the
tachyonic soliton solution on a D-brane, representing a D-brane of lower
dimension, but also noted that there are other solutions with no obvious
interpretation as D-branes. In this section we shall argue that all these
are equivalent solutions, in the same sense that different values of
$\theta$ and $\phi$ for $r=0$ represent the same configuration of a
particle moving in 3-dimensions.

The configuration we study is the same one studied in ref.\cite{0007226},
$-$ a
D-25-brane in the presence of constant background antisymmetric tensor
field $B_{\mu\nu}$.
As in \cite{0007226} we take space-time to be Euclidean, and take $B$ to
be of
rank 26. As shown in \cite{9908142,9908019,9912070}, this system has many
different
descriptions
with different non-commutativity parameters $\Theta^{\mu\nu}$. These
different
descriptions are labelled by an anti-symmetric tensor $\Phi_{\mu\nu}$, in
terms of which $\Theta^{\mu\nu}$, the open string metric $G_{\mu\nu}$ and
the open
string coupling constant $G_o$ are given by:
\ben \label{e3.1a}
(G+2\pi\alpha'\Phi)^{-1} = -{\Theta\over 2\pi\alpha'} + (g+2\pi\alpha'
B)^{-1}\, , \nonumber \\
G_o^2 = g_c \bigg({\det(G+2\pi\alpha' \Phi)\over \det(g+2\pi\alpha'
B)}\bigg)^{1/2}\, .
\een
Here $g_{\mu\nu}$ is the closed string metric and $g_c$ is the closed
string
coupling constant. One convenient choice of $\Phi$ is:
\be \label{e3.1b}
\Phi^{(1)}=0\, ,
\ee
with the corresponding non-commutativity parameter $\Theta_{(1)}$ and the open
string metric and coupling constants $G_{(1)}$ and $G_{o(1)}$ determined from
eq.\refb{e3.1a}. The solutions given in \cite{0007226} were constructed
in these variables. The other choice, which is useful in constructing
background independent variables\cite{0007226,0008013}, is
\be \label{e3.1c}
\Phi^{(2)}=-B\, ,
\ee
which gives
\be \label{e3.1d}
\Theta_{(2)}=B^{-1}, \qquad G_{(2)}=-(2\pi\alpha')^2 B g^{-1} B, 
\qquad G_{o(2)}^2 =
g_c det(2\pi\alpha' B g^{-1})^{1/2}\, .
\ee
We shall denote by $\wh A^{(1)}_\mu$ and $\wh A^{(2)}_\mu$ 
the non-commutative gauge
fields in the first and the second description respectively, and by 
\be \label{e3.1e}
\wh F^{(s)}_{\mu\nu} = \p_\mu \wh A^{(s)}_\nu - 
\p_\nu \wh A^{(s)}_\mu - i [\wh A^{(s)}_\mu, \wh
A^{(s)}_\nu]_{\Theta_{(s)}}, \qquad s=1,2
\ee
the corresponding non-commutative
field strength.\footnote{We shall use real coordinate
system $x^\mu$ instead of the
complex
one used in \cite{0007226}. Also, for gauge fields we use the sign
convention
of 
ref.\cite{9908142} which differs from that of ref.\cite{0007226} by a
minus
sign.} Here
$[,]_{\Theta_{(s)}}$ denotes that we should compute the commutator using the 
non-commutative product with parameter $\Theta_{(s)}$:
\be \label{encom}
[x^\mu, x^\nu]_{\Theta_{(s)}} = i \Theta_{(s)}^{\mu\nu}\, .
\ee
{}From now on, we shall
drop the
subscript $\Theta_{(s)}$ 
on various commutators. Defining new variables, 
\be
\label{e3.1f}
C_\mu^{(s)}=(\Theta_{(s)}^{-1})_{\mu\nu} x^\nu + \wh A^{(s)}_\mu\, ,
\ee
we get 
\be \label{e3.1g}
\wh F^{(s)}_{\mu\nu} = -i[C_\mu^{(s)}, C_\nu^{(s)}] 
+ (\Theta_{(s)}^{-1})_{\mu\nu}\, .
\ee
The effective action involving the tachyon and the gauge field in the two
descriptions is given by\cite{9908142,0007226,0008013}:
\ben \label{e3.2}
&& {\sqrt{\det\Theta_{(s)}}\over G_{o(s)}^2\alpha^{\prime 13}
(2\pi)^{12}}
Tr\bigg[\VV(T)
\sqrt{\det\big((G_{(s)})_{\mu\nu}+2\pi\alpha'\Phi^{(s)}_{\mu\nu}
-2\pi\alpha'(i[C^{(s)}_\mu,C^{(s)}_\nu]
-
(\Theta^{-1})_{\mu\nu})\big)} \nonumber \\
&& + \alpha' f(T) [C^{(s)}_\mu, T] [T, C^{(s)}_\nu] 
(G_{(s)})^{\mu\nu} \sqrt{\det (G_{(s)})} +\ldots
\bigg]\, ,
\een
where $\VV(T)$ is the tachyon potential which vanishes at $T=T_{min}$, and
$f(T)$ is an unknown function of $T$.
Here $Tr$ denotes trace over infinite dimensional matrices used to
represent various functions of $x^\mu$ following the procedure of
\cite{0003160}. As is customary, we shall denote by $N$ the dimension of
these
matrices, with the understanding that $N$ is actually infinite. $\ldots$
in eq.\refb{e3.2} denote various higher derivative terms.

The relationship between the two sets of variables can be found by noting
that the ordinary gauge field strength $F_{\mu\nu}$ is related to $\wh
F^{(s)}_{\mu\nu}$ through the relation:
\be \label{e3.3}
\wh F^{(s)} = (1 + F\Theta_{(s)})^{-1} F, \qquad
F = \wh F^{(s)} (1 - \Theta_{(s)} \wh F^{(s)})^{-1}, \qquad \hbox{for} 
\quad s=1,2\, .
\ee
In particular, if $[C^{(s)}_\mu, C^{(s)}_\nu]=0$, 
it corresponds to $F=\infty$ for
both values of $s$. Since under an $U(N)$ gauge
transformation
\be \label{euinf}
C^{(s)}_\mu\to U C^{(s)}_\mu U^\dagger\, ,
\ee
$C^{(1)}_\mu=0$ ($C^{(2)}_\mu=0$) represents $U(N)$ invariant gauge field
configurations\cite{0007226}.

For the choice $s=2$, one can also define the background independent
variables $X^\mu$ as follows:
\be \label{e3.4}
X^\mu=(\Theta_{(2)})^{\mu\nu}C^{(2)}_\nu\, .
\ee
The action expressed in terms of these variables takes the
form\cite{0007226,0008013}:
\be \label{e3.5}
{1\over g_c \alpha^{\prime 13} (2\pi)^{12}}
Tr\bigg[\VV(T)
\sqrt{\det\big(\delta_\mu^{~\nu}-2\pi\alpha'ig_{\mu\rho}[X^\rho,X^\nu]
\big)} + \alpha' f(T) g_{\mu\nu} [X^\mu, T] [T, X^\nu]
+\ldots
\bigg]\, ,
\ee
Written in these variables, the action does not depend on the choice of
the background $B$ field.

Denoting by $I_m$ the $m\times m$ identity matrix for any integer $m$, the
tachyonic vacuum corresponds to $T=T_{min}I_N$, and $\VV(T)$ vanishes 
in this vacuum. This
implies that the kinetic term
for the $C^{(s)}_\mu$ fields vanishes at 
$T=T_{min} I_N$ and hence the $C^{(s)}_\mu$'s
themselves are not good coordinates around this point.\footnote{As 
we lack sufficient intuition about theory
of matrices, it is hard to conclude just by looking at eqs.\refb{e3.2} or
\refb{e3.5} that the variables appearing in these equations are wrong
variables. Here we are using our intuition from conventional field
theories; $-$ since when expressed in the space-time language the kinetic
term for the gauge fields vanish, different gauge field
configurations, labelled by different values of $C_\mu$ ($X^\mu$), should
correspond to same physical configurations at $T=T_{min}I_N$.} 
Instead we should
choose new variables
\be \label{e3.6}
\wt C_\mu = h_\mu(T, \vec C)\, , \qquad \wt T  = h(T, \vec C)
\ee
to describe the field configuration around this point. (Note that we have
dropped the superscript $(s)$ from $C$, $-$ the discussion below holds for
either choice of $s$, and also for $\vec C$ replaced by the background
independent variables $\vec X$. The precise form of the functions $h_\mu$ 
and $h$ will of course depend on whether we choose $C^{(1)}_\mu$,
$C^{(2)}_\mu$ or $X^\mu$ in the arguments of these functions.) Vanishing
of the $C_\mu$ kinetic term at $T=T_{min}I_N$
suggests that the functions $h_\mu$ and $h$ have the
property:\footnote{The important point here is that $h_\mu(T_{min} I_N,
\vec C)$ and $h(T_{min}I_N, \vec C)$ should be independent of $\vec C$.
A constant shift in the definition of $\wt C_\mu$ can then be used to set
$h_\mu(T_{min}I_N, \vec C)$ to zero.}$^,$\footnote{For our purpose it
will be sufficient if these relations hold for those $C_\mu$'s which
satisfy equations of motion.}
\be \label{e3.7}
h_\mu(T_{min} I_N, \vec C) = 0\, , 
\qquad h(T_{min} I_N, \vec C) = \wt T_{min} I_N\, ,
\ee
where $\wt T_{min}$ is some constant which could also be set to zero by
redefining $\wt T$.
We shall assume that the functions $h_\mu$ and $h$ appearing in
eq.\refb{e3.6} may be expressed as (infinite) sum of products of various
powers (possibly fractional) of $T$ and $C_\mu$, so that eqs.\refb{e3.7}
hold as operator
equations irrespective of the dimension of the operators. In that case, for
any non-singular matrix $S$, we have
\be \label{e3.8}
h_\mu(S T S^{-1}, S \vec CS^{-1}) = S h_\mu(T, \vec C) S^{-1}\, , \qquad
h(S T S^{-1}, S \vec CS^{-1}) = S h(T, \vec C) S^{-1}\, .
\ee
In particular, if we take $S$ to be an $N\times N$ unitary matrix, then
the above relation shows that $\wt T$ and $\wt C_\mu$ defined in
eq.\refb{e3.6} transform in the adjoint representation of $U(N)$. Thus we
can get an $U(N)$ invariant configuration by taking:
\be \label{e3.9}
\wt T=\wt T_{min} I_N, \qquad \wt C_\mu = 0\, .
\ee
We identify this as the `nothing' state. Note however, that this does not
require us to take $C_\mu=0$; any finite $C_\mu$ corresponds to this
vacuum
when $T=T_{min} I_N$. Thus {\it the coordinate redefinition suggested in
eqs.\refb{e3.6}, \refb{e3.7} gets rid of the problem of having degenerate
vacua labelled by different values of the gauge field vacuum
expectation values.}

The direct evidence for eq.\refb{e3.7} comes from examining eq.\refb{e3.6}
with
$\vec C$ replaced by $\vec X$. If we consider a configuration of commuting
$X^\mu$'s, then the eigenvalues of $X^\mu$ can be 
regarded as the positions of
D-instantons making up the original
D-25-brane\cite{CORNALBA,9909176,0008013}. 
Now we can
invoke
the
string field theory analysis of ref.\cite{0007153,0008033} to argue that
when the tachyon
rolls down to its minimum $T_{min}$, there is no flat direction of the
potential, and hence all the different (commuting) values of $X^\mu$ must
correspond to the same point in the configuration space.
Eqs.\refb{e3.6}, \refb{e3.7}
(with $\vec C$ replaced by $\vec X$) clearly incorporates this. 

Let us now turn to the soliton solutions. Expressed in terms of the
original variables $T$, $C_\mu$, these solutions are typically of
the block diagonal form\cite{0007226}:
\be \label{e3.10}
T = \pmatrix{T_{max} I_M & \cr & T_{min} I_{N-M}}, \qquad
C_\mu = \pmatrix{ S_\mu & \cr & V_\mu}\, ,
\ee
where $T_{max}$ is the value of the tachyon at which the potential has a
local maximum, representing the original D-brane before tachyon
condensation, $M$ is an integer (possibly infinite) and $S_\mu$ and $V_\mu$
are $M\times M$ and $(N-M)\times (N-M)$ matrices respectively. The
unwanted degeneracy of the solutions comes from different choices of
$V_\mu$.
$S_\mu$'s on the other hand represent world-volume fields on the
resulting solitonic brane, and different values of $S_\mu$ correspond to
different background field configurations on this brane. 

Using \refb{e3.6}, and the assumption that $h_\mu$ and $h$ involve
sum of products of various powers of $T$ and $C_\mu$,
the solution \refb{e3.10} can be
rewritten as
\be \label{e3.11}
\wt C_\mu = \pmatrix{ h_\mu(T_{max} I_{M}, \vec S) & \cr & h_\mu(T_{min}
I_{N-M}, \vec V)}\, , \qquad
\wt T = \pmatrix{ h(T_{max} I_{M}, \vec S) & \cr & h(T_{min}
I_{N-M}, \vec V)}\, .
\ee
Using eq.\refb{e3.7}, and the fact that it represents operator
relations irrespective  of the dimension of the operators, we can
rewrite eq.\refb{e3.11} as
\be \label{e3.12}
\wt C_\mu = \pmatrix{ h_\mu(T_{max} I_{M}, \vec S) & \cr & 0 }\, , \qquad
\wt T = \pmatrix{ h(T_{max} I_{M}, \vec S) & \cr & \wt T_{min}
I_{N-M}}\, .
\ee
{}From this we see that in terms of the new variables solutions with 
different background values of $\vec V$ correspond to the same field
configuration. Thus {\it the apparent unwanted degeneracy of
the
soliton solutions disappears when we make the right
choice of variables in
describing these solutions}.

Since different background values of $V_\mu$ correspond to
the same tachyonic soliton configuration, we can choose
any of these representative values to analyse the tachyon condensation
problem. As was shown in refs.\cite{0005006,0005031,0007226}, many exact
results can be obtained
by taking $\vec V$ to be zero, which corresponds to taking the
value of the ordinary magnetic field strength away from the location of
the soliton to infinity. Thus if our
proposal, $-$ that the
correct choice of coordinates is given by $\wt C_\mu$ and $\wt T$ defined in
eqs.\refb{e3.6}, \refb{e3.7} $-$ is correct, then the exact results for
$\vec V=0$ can be used to conclude that the same results also hold for
$\vec V\ne 0$. Translated to conventional language, this will imply
that the exact results, obtained in the limit of large asymptotic magnetic
field on the original D-25 brane, are also valid when the asymptotic
magnetic field is finite.\footnote{A similar remark was made in
ref.\cite{0005031}.} 
For magnetic field in the plane transverse to the soliton this is a
surprising result, since
for finite
asymptotic magnetic field higher derivative terms in the effective action 
can no longer be ignored
in the analysis, and the shape of the soliton is expected to change. But
if
eqs.\refb{e3.6}, \refb{e3.7} are
correct, then this must be a gauge artifact.

An indirect evidence for this follows from the fact that the D-23 brane,
after all, is described by a specific conformal field theory, and hence
must be described by a unique configuration in string theory. Thus any two
different configurations, describing the D-23-brane in the same closed
string background, must be gauge equivalent. 

\sectiono{The Fate of the Bifundamental States} \label{s4}

The analysis of refs.\cite{0005031,0007226} found the existence of certain
finite mass
open string states on the world-volume of the soliton which transform in
the bifundamental representation of $U(M)\times U(N-M)$. These arise from
fluctuations of off block-diagonal components of $C_\mu$ and $T$ around
the background \refb{e3.11}. These states are
not present in the excitation spectrum of a
lower dimensional D-brane. Hence the existence of these states on the
soliton world-volume poses a problem for identifying the solitons as lower
dimensional D-branes.

Physically, these modes
represent open string states with one end on
the D-brane soliton and the other end on the tachyonic
vacuum\cite{0006071}. Since open
strings cannot end on the vacuum, these states must be absent from the
spectrum. A possible explanation for the absence of these states was given
in \cite{9909062} (for related but alternative explanations, see
refs.\cite{9901159,0002223}).
The main point here is that since the kinetic term for the U(1) gauge
field living in the vacuum vanishes (we are using the original variables
before the field redefinition), the equations of motion of the gauge field
will force the currents coupled to the gauge field to vanish. These
constraints will remove from the spectrum the states which are charged
under the gauge field. This includes the open string states with one end
on the D-brane soliton and the other end in the tachyonic vacuum.

The existence of these constraints can be seen explicitly in the present
framework. Since we have argued that backgrounds labelled by different
choices of $V_\mu$ in eq.\refb{e3.10} correspond to physically identical
configurations, we shall choose a background where $V_\mu$ are
non-vanishing, giving finite
asymptotic magnetic field. We can now consider
fluctuating fields of the form:
\be \label{e4.1}
T = \pmatrix{T_{max} I_M & \cr & T_{min} I_{N-M}} + \pmatrix{\wh S & \wh
W\cr \wh W^\dagger & \wh V}, \qquad
C_\mu = \pmatrix{ S_\mu & \cr & V_\mu} + \pmatrix{\wh S_\mu & \wh W_\mu\cr
\wh W_\mu^\dagger & \wh V_\mu}\, ,
\ee
where the hatted variables correspond to fluctuations around the
background \refb{e3.10}.
In particular, the fluctuations $\wh V_\mu$ represent fluctuations of the
gauge
fields in the vacuum {\it outside the soliton}. Since $\VV(T_{min}
I_{N-M})=0$, we see
from eq.\refb{e3.2} that the regular kinetic term for $\wh V_\mu$ (which
would have been proportional to $Tr[V_\mu,\wh V_\nu][V^\mu,\wh V^\nu]$ had
$\VV(T_{min}I_{N-M})$ been finite) vanishes
in the background \refb{e3.10}. Thus the equations of motion of $\wh V_\mu$
does not contain any term linear in $\wh V_\mu$, and hence, instead of
determining the gauge fields in terms of the currents, imposes constraints
containing quadratic and higher powers of the
bifundamental fields $\wh W$, $\wh W_\mu$. These
constrains ensure absence of states charged under $\wh V_\mu$.

\noindent{Acknowledgement}: I would like to thank R. Gopakumar, S.
Minwalla and A. Strominger for many useful discussions during various
stages of this work. I would also like to thank R. Gopakumar, S.
Minwalla and B. Zwiebach for careful reading of the manuscript and for
suggestions for
improvement.

\end{document}